\documentclass[conference]{IEEEtran}
\IEEEoverridecommandlockouts
\usepackage{cite}
\usepackage{amsmath,amssymb,amsfonts}
\usepackage{algorithmic}
\usepackage{graphicx}
\usepackage{textcomp}
\usepackage{xcolor}

\usepackage[utf8]{inputenc}
\usepackage[most]{tcolorbox}

\usepackage{booktabs}

\usepackage{graphicx}
\usepackage{multirow}
\usepackage{cite} 
\usepackage[hidelinks,bookmarks=false]{hyperref}

\usepackage{pgfplots}
\usepackage{pgfplotstable}
\usepgfplotslibrary{statistics}
\usepackage{cleveref}

\makeatletter
\let\MYcaption\@makecaption
\makeatother
\usepackage[font=footnotesize]{subcaption}
\makeatletter
\let\@makecaption\MYcaption
\makeatother

\usepackage{comment}
\usepackage{listings}

\usepackage{circledsteps}

\usetikzlibrary{patterns}

\usepackage{blindtext}

\usepackage{acronym}

\usepackage{xcolor, colortbl}

\usepackage{float}
\usepackage{textcomp}
\definecolor{linkcolor}{RGB}{219, 48, 122}

\usepackage{algorithm}
\usepackage{algorithmic}
\usepackage{url}
\usepackage[most]{tcolorbox}
\usepackage{varwidth}
\tcbuselibrary{xparse}
\tcbuselibrary{skins}
\usepackage[all=normal, floats=tight,indent=tight]{savetrees}
\usepackage{pifont}
\usepackage{framed}
\usepackage{soul}
\usepackage{multirow}
\usepackage{array}
\newcolumntype{L}[1]{>{\raggedright\let\newline\\\arraybackslash\hspace{0pt}}m{#1}}
\newcolumntype{C}[1]{>{\centering\let\newline\\\arraybackslash\hspace{0pt}}m{#1}}
\newcolumntype{R}[1]{>{\raggedleft\let\newline\\\arraybackslash\hspace{0pt}}m{#1}}
\newcolumntype{H}{>{\collectcell\lstinline}l<{\endcollectcell}}
\usepackage{url}

\newcommand{\sol}{LASHED}

\pgfplotsset{compat=1.18}
\usepackage[T1]{fontenc}

\input{Sections/preamble/preamblelistingsconf}
\acrodef{CPS}{Cyber-Physical System}
\acrodef{IoT}{Internet of Things}
\acrodef{HDL}{Hardware Description Language}
\acrodef{RTL}{Register-Transfer Level}
\acrodef{CAD}{Computer-Aided Design}
\acrodef{EDA}{Electronic Design Automation}
\acrodef{HPC}{High-Performance Computing}
\acrodef{DL}{deep learning}
\acrodef{ML}{machine learning}
\acrodef{NLP}{natural language processing}
\acrodef{IC}{Integrated Circuit}
\acrodef{CWE}[CWE]{Common Weakness Enumeration}
\acrodef{CVE}[CVE]{Common Vulnerabilities and Exposures}
\acrodef{LLM}[LLM]{large language model}
\acrodef{NMT}[NMT]{neural machine translation}
\newcommand{\ignore}[1]{{}}

\newcommand{\squishlist}{
	\begin{list}{$\bullet$}
		{ \setlength{\itemsep}{0pt}
			\setlength{\parsep}{1pt}
			\setlength{\topsep}{1pt}
			\setlength{\partopsep}{0pt}
			\setlength{\leftmargin}{0.9em}
			\setlength{\labelwidth}{1.5em}
			\setlength{\labelsep}{0.4em} } }
	\newcommand{\squishend}{
	\end{list}  }

\definecolor{graphFirst}{RGB}{2,136,209} 
\definecolor{graphSecond}{RGB}{211,47,47} 
\definecolor{graphThird}{RGB}{245,124,0} 
\definecolor{graphFourth}{RGB}{56,142,60} 
\definecolor{graphFifth}{RGB}{81,45,168} 
\definecolor{graphSixth}{RGB}{69,90,100} 
\definecolor{graphSeventh}{RGB}{251,192,45} 
\definecolor{backgroundSecond}{RGB}{239,154,154} 
\definecolor{backgroundThird}{RGB}{255,204,128} 
\definecolor{backgroundFourth}{RGB}{165,214,167} 
\definecolor{backgroundFifth}{RGB}{179,157,219} 
\definecolor{backgroundSixth}{RGB}{176,190,197} 
\definecolor{backgroundSeventh}{RGB}{255,245,157} 

\def\BibTeX{{\rm B\kern-.05em{\sc i\kern-.025em b}\kern-.08em
    T\kern-.1667em\lower.7ex\hbox{E}\kern-.125emX}}

\makeatletter
\def\endthebibliography{%
  \def\@noitemerr{\@latex@warning{Empty `thebibliography' environment}}%
  \endlist
}
\makeatother

\pdfoutput=1 

\begin{document}

\title{LASHED: \underline{L}LMs \underline{A}nd \underline{S}tatic \underline{H}ardware Analysis for \underline{E}arly \underline{D}etection of RTL Bugs\\
}


\author{
\IEEEauthorblockN{Baleegh Ahmad}
\IEEEauthorblockA{
\textit{NYU Tandon}\\
Brooklyn, USA \\
ba1283@nyu.edu}
\and
\IEEEauthorblockN{Hammond Pearce}
\IEEEauthorblockA{
\textit{University of New South Wales}\\
Sydney, Australia \\
hammond.pearce@unsw.edu.au}
\and
\IEEEauthorblockN{Ramesh Karri}
\IEEEauthorblockA{
\textit{NYU Tandon}\\
Brooklyn, USA \\
rkarri@nyu.edu}
\and
\IEEEauthorblockN{Benjamin Tan}
\IEEEauthorblockA{
\textit{University of Calgary}\\
Calgary, Canada \\
benjamin.tan1@ucalgary.ca}
}

\maketitle

\thispagestyle{plain}
\pagestyle{plain}

\begin{abstract}
While static analysis is useful in detecting early-stage hardware security bugs, its efficacy is limited because it requires information to form checks and is often unable to explain the security impact of a detected vulnerability.
Large Language Models can be useful in filling these gaps by identifying relevant assets, removing false violations flagged by static analysis tools, and explaining the reported violations.
\sol{} combines the two approaches (LLMs and Static Analysis) to overcome each other's limitations for hardware security bug detection.
We investigate our approach on four open-source SoCs for five Common Weakness Enumerations (CWEs) and present strategies for improvement with better prompt engineering. We find that 87.5\% of instances flagged by our recommended scheme are plausible CWEs. In-context learning and asking the model to `think again' improves \sol{}'s precision.
\end{abstract}


\acresetall

\section{Introduction\label{sec:intro}}

Security vulnerabilities in hardware are difficult to detect~\cite{dessouky_hardfails_2019}. It is critical to identify them early on in the system-on-chip (SoC) design life-cycle because of the higher costs of fixing issues downstream (pre-silicon) or even recalls (post-silicon)~\cite{mitra_post-silicon_2010}. An exhaustive search for these defects is not possible because of the high complexity of modern processors and SoCs. Therefore, there is a need for innovative solutions that provide \textbf{early-stage} information on potential security issues at the \ac{RTL}. 

Existing strategies for security verification include simulation with test benches, formal assertions~\cite{ray_invited_2019,he_soc_2019}, hardware fuzzing~\cite{hur_difuzzrtl_2021,trippel_fuzzing_2022} and information flow tracking~\cite{ardeshiricham_register_2017,hu_hardware_2021}.
Security checks ``as-you-go,'' while implementation is ongoing, are more challenging. 
Recent works have proposed static analysis~\cite{ahmad_dont_2022,bidmeshki_hunting_2021} and \acp{LLM}~\cite{ahmad_flag_2023, tarek_socurellm_2024, akyash_self-hwdebug_2024} for this purpose. 
Static analysis checks source code 
without ``executing'' it, 
e.g., without simulating the design to perform directed tests. 
The code is instead checked against 
a set of coding patterns that can indicate undesirable behavior.
Linters \cite{vc_spyglass_lint_synopsys_2022,jasperlint_jasper_2022} and formal verification tools~\cite{noauthor_vc_2022,cadence_jasper_2022} are the two most commonly used static analysis~\cite{ahmad_dont_2022, hansson_continuous_2014} methods for RTL. 
Linting is the automated checking of source code for stylistic, structural, design and programmatic checks~\cite{mcnutt_linting_2018}, and can 
include data-flow analysis and control-flow analysis, as well as more abstract techniques such as pattern matching for bug-specific heuristic patterns.
Formal methods use mathematical models to analyze and verify a design
~\cite{woodcock_formal_2009}. 

Development and use of \acp{LLM} 
~\cite{chen_evaluating_2021} has provided a possible means to detect bugs in code without the explicit need for a fully mature testing framework.
LLMs have been used for RTL generation~\cite{thakur_verigen_2024} and repair~\cite{ahmad_hardware_2024} with reasonable degrees of success, but their ability to \textbf{detect} security bugs has yet to be proven. 
In part, this is because \acp{LLM} do not verify their outputs. Static analysis can address this shortcoming. 
Prior work in the software space has explored the combination of LLMs and static analysis; for example, IRIS~\cite{li_llm-assisted_2024} uses CodeQL as the static analysis tool coupled with LLMs to detect the code injection vulnerabilities in Java code. Chapman et al. present an approach to interleave LLMs with the EESI static analysis tool to detect the issue of error-specific inference~\cite{chapman_interleaving_2024}. LLIFT~\cite{li_enhancing_2024} uses the LLM and static analysis combination to detect use before initialization bugs within the Linux kernel. 

Taking inspiration from such works, we propose a strategy which uses LLMs and Static Analysis together, i.e., \sol{}. 
We use hardware \acp{CWE}~\cite{the_mitre_corporation_mitre_cwe_2022}, which provide examples of vulnerability categories to aid the generalizability of our approach. 
We use the LLM for three tasks: i) identifying security relevant assets, ii) removing false positives from static analysis, and iii) explaining the security issue posed by a reported violation.
To identify security assets, the LLM uses the RTL source code and Hardware \acp{CWE}. 
The hardware static analysis tools formulate checks that could indicate the presence of certain CWEs. 
We use either linting or formal property verification for static analysis, depending on the nature of the CWE. 
The linting violations and failing assertions are returned to the LLM to prune out those that do not pose any security threat and provide explanations for the ones that do. 
Our contributions are: 
\begin{itemize}
    \item We present the first framework that combines Static Analysis and LLMs to detect security issues in RTL code. The details of this tool are described in~\autoref{sec:lashed}.
    \item We validate LASHED on four open source SoCs, described in~\autoref{sec:experiments}. Results are presented in~\autoref{sec:results}.
    \item We investigate the impact of in-context learning and prompting to insist that the LLM reason through its assessments. The outcomes are analyzed in~\autoref{subsec:analysis}.
\end{itemize}

\section{LASHED\label{sec:lashed}}

\begin{figure*}[t]
    \centering
    \includegraphics[width=0.95\linewidth]{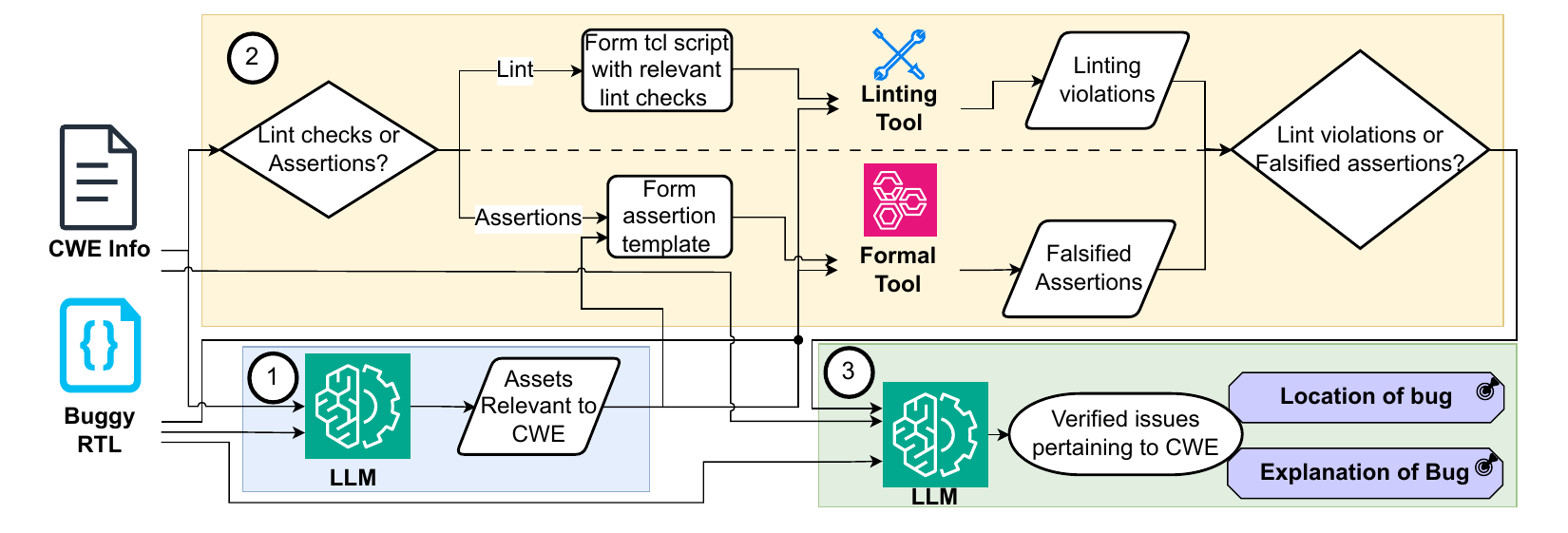}
    \caption{\sol{} framework.  \textbf{1) Assets Identification (AssetID):} LLM identifies assets relevant to a CWE in the RTL. \textbf{2) Static Analysis:} Depending on CWE, a linting or assertion-based strategy is used. A linter or formal property verification tool checks for the lint tags identified or assertions formed. \textbf{3) Contextualization:} LLM reasons and prunes linting violations or falsified properties. 
    LLM gives an explanation if there is a security issue.}
    \label{fig:framework}
    
\end{figure*}

\sol{} takes RTL source and CWE information as inputs and outputs potential issues pertaining to the CWE. The CWEs covered are detailed in~\autoref{subsec:cwes}. The output contains the bug code, its explanation, and its location.
The framework for our approach is shown in~\autoref{fig:framework}.
It can be broken down into three main steps, i.e., \textbf{Assets Identification}, \textbf{Static Analysis}, and \textbf{Contextualization}.
We illustrate these steps through two motivating examples taken from the Hack@DAC 2021 buggy OpenPiton SoC shown in~\autoref{fig:motivating-example-cwe1191} and~\autoref{fig:motivating-example-cwe1233}.

The vulnerability in~\autoref{fig:motivating-example-cwe1191-srccode} shows an instance of CWE 1191 where only the least
significant 32 bits of the secret message are used to authenticate the JTAG access control module. This makes the access control susceptible to brute-force attacks.
The vulnerability in~\autoref{fig:motivating-example-cwe1233-srccode} shows an instance of CWE 1233 where security-sensitive registers are missing lock bit protection in the Direct Memory Access (DMA) wrapper module. An adversary can modify them from software.

\begin{figure}[h!]
\centering

\begin{subfigure}[b]{0.95\linewidth}
\begin{lstlisting}[language=Verilog, frame=lines, escapeinside={(*}{*)}, linebackgroundcolor={\ifnum\value{lstnumber}>1
                \ifnum\value{lstnumber}<3
                    \color{pink}
                \fi
            \fi        }]
if (pass_mode) begin
    pass_data = { {60{8'h00}}, data_d};
    state_d = PassChk;
    pass_mode = 1'b0;
\end{lstlisting}
\caption{CWE 1191 in JTAG access control mechanism. Only the least significant 32 bits of the secret message are used for authentication.}
\label{fig:motivating-example-cwe1191-srccode}
\end{subfigure}

\begin{subfigure}[b]{0.95\linewidth}
\vspace{3mm}
\begin{lstlisting}[language=json, numbers=none, frame=lines ]
"access_control_related_signals": [ "pass_data", "pass_hash", "exp_hash", "hmac_ready", "data_d" ...],
\end{lstlisting}
\caption{ Assets identification using gpt-4o-mini, prompt variation v0.}
\label{fig:motivating-example-cwe1191-assets-identification}
\end{subfigure}

\begin{subfigure}[b]{0.95\linewidth}
\vspace{3mm}
\begin{lstlisting}[language=json, numbers=none, frame=lines ]
"relevant_violations": [{
    "line_no": 158,
    "statement": "pass_data = {{60{8'b0}},data_d} ; ",
    "lhsexpr": "pass_data",
    "security_sensitive_signal": "pass_data" }]
\end{lstlisting}
\caption{ Lint violation using VC SpyGlass Lint and Verific.}
\label{fig:motivating-example-cwe1191-sa-outputs}
\end{subfigure}

\begin{subfigure}[b]{0.95\linewidth}
\vspace{3mm}
\begin{lstlisting}[language=json, numbers=none, frame=lines ]
"insecure": true,
"explanation": The assignment to 'pass_data' defines a 256-bit wide data signal (with 60 zeros prepended to 'data_d'), ... If unauthorized access is not controlled before 'pass_data' is evaluated or used, an attacker could send malicious data to the device and compromise intended security operations, making it a potential violation of proper access control in line with CWE 1191.


\end{lstlisting}
\caption{ Contextualization using gpt-4o-mini variation v0.}
\label{fig:motivating-example-cwe1191-contextualization}
\end{subfigure}

\caption{Motivating example for CWE 1191 to illustrate LASHED flow for linter-based strategy.}
\label{fig:motivating-example-cwe1191}

\end{figure}

\begin{figure}[h!]
\centering

\begin{subfigure}[b]{0.95\linewidth}
\vspace{3mm}
\begin{lstlisting}[language=Verilog, frame=lines, escapeinside={(*}{*)}, linebackgroundcolor={\ifnum\value{lstnumber}>5
                \ifnum\value{lstnumber}<10
                    \color{pink}
                \fi
            \fi        }]
input logic [7 :0] reglk_ctrl_i; // register lock values
...
    else if(en && we)
        case(address[7:3])
            0: start_reg <= wdata;
            ...
            8: end_reg <= wdata;
\end{lstlisting}
\caption{CWE 1233 in Direct Memory Access wrapper. Security sensitive registers are missing lock bit protection.}
\label{fig:motivating-example-cwe1233-srccode}

\end{subfigure}

\begin{subfigure}[b]{0.95\linewidth}
\vspace{3mm}
\begin{lstlisting}[language=json, numbers=none, frame=lines ]
"relevant-signals": {
    "security_sensitive_signals_info": [
        {"lock_signal": "reglk_ctrl_i[0]",
         "security_sensitive_signal": "start_reg"},
        ...
        {"lock_signal": "reglk_ctrl_i[7]",
         "security_sensitive_signal": "core_lock_reg"},
    "reset_conditions": "~(rst_ni && ~rst_8)", ...}
\end{lstlisting}
\caption{ Assets identification using gpt-4o-mini, prompt variation v0.}
\label{fig:motivating-example-cwe1233-assets-identification}
\end{subfigure}

\begin{subfigure}[b]{0.95\linewidth}
\vspace{3mm}
\begin{lstlisting}[language=Verilog, numbers=none, frame=lines ]
@(posedge clk_i)
disable iff (~(rst_ni && ~rst_8)) reglk_ctrl_i[7] == '1 |=> $stable(core_lock_reg);
\end{lstlisting}
\caption{ Assertion formation for \texttt{core\_lock\_reg} signal in DMA wrapper using VC Formal Property Verification. This assertion was falsified.}
\label{fig:motivating-example-cwe1233-sa-outputs}
\end{subfigure}

\begin{subfigure}[b]{0.95\linewidth}
\vspace{3mm}
\begin{lstlisting}[language=json, numbers=none, frame=lines ]
"insecure": true,
"explanation": The signal 'core_lock_reg' is managed but can be set to zero inadvertently, thereby allowing modification of the system's important registers. It lacks the necessary stability to prevent unauthorized access.

\end{lstlisting}
\caption{ Contextualization using gpt-4o-mini variation v0.}
\label{fig:motivating-example-cwe1233-contextualization}
\end{subfigure}

\caption{Motivating example for CWE 1233 to illustrate LASHED flow for assertion based strategy.}
\label{fig:motivating-example-cwe1233}

\end{figure}

\subsection{Assets Identification (AssetID)}

We use an LLM to identify assets relevant to the CWE that \sol{} is scanning for. 
We prompt the LLM with a system prompt that primes it as a hardware security expert searching for the CWE. 
This is followed by the description of the CWE  from the MITRE website. 
There are numerous ways to structure the prompt and alter the included information which we investigate (see Section~\ref{subsec:pe-variations}). For instance, for prompt strategy \textit{v1} the system prompt is followed by an example of the CWE in Verilog code derived from MITRE website, including identifying relevant assets and explaining why this example poses the CWE. 
After priming the LLM with the system prompt, the LLM is given a user prompt that contains the RTL source and instructions on what kinds of signals are to be identified. The identified assets are sent for static analysis.
The components of the prompt are illustrated below:

\begin{lstlisting}[language=Python, numbers=none, frame=lines ]
system_prompt = "You are a hardware security expert. Your task is to analyze Verilog code for potential CWE-<x> bugs. CWE-<x> is <description of CWE>."

user_prompt = "What are the <relevant signals>? <Typical nature of such signals>" + <RTL source code>.

if variation == 'v1':
    user_prompt = "<example of CWE in RTL> + <explanation of security issue> + <assets identified in this case>" + user_prompt
\end{lstlisting}

Asset identification of the motivating examples is shown in~\autoref{fig:motivating-example-cwe1191-assets-identification} and~\autoref{fig:motivating-example-cwe1233-assets-identification}.
The relevant access control signals for CWE-1191  example \texttt{\{pass\_data,data\_d\}} are correctly identified along with unrelated signals.  
For CWE-1233, the relevant security sensitive signals \texttt{\{start\_reg ... core\_lock\_reg\}} and their ``expected'' lock signals \texttt{\{reglk\_ctrl[*]\}} are identified. The reset conditions are also identified for assertion formation in the next step.

\subsection{Static Analysis}
Depending on the CWE, we decide whether a linting-based or assertion-based strategy is more appropriate. 
If the CWE is typically present alongside some ``structural'' or ``coding style'' imperfections, we use lint checks. This is the case for CWEs 1191 and 1300. 
Conversely, if the CWE requires verifying whether a signal behaves appropriately depending on the value(s) of other signal(s) or if one signal flows to another, we use assertions. This applies to CWEs 1231, 1233 and 1244. 

\subsubsection{Linter-based}
We select relevant checks from the $\sim$1000 VC SpyGlass Lint~\cite{vc_spyglass_lint_synopsys_2022} tags in the functional lint tag database 
based on our understanding of the CWE.
For some CWEs, we develop custom lint checks using Verific~\cite{verific_verific_2022}, where, after obtaining the Abstract Syntax Tree (AST) of any module, we traverse it to check for some structural or stylistic element.
If the checks' results include the assets identified previously, the results are considered \textit{violations}. 
These are sent to the next step, i.e., contextualization.

For CWE-1191, improper access control for debug occurs when a signal containing the user input (password) is not properly assigned a value. 
We check the code for the following: \textit{[Width mismatch, Reverse Connected busses, Improper range index, Concatenation in array assign, Concatenation using unsized numbers, RHS has concatenation]} -- if the assignment contains the previously identified access control signals, the signal and assignment are reported as a violation.
\autoref{fig:motivating-example-cwe1191-sa-outputs} shows reported violations for the CWE-1191 example. 

For CWE-1300, we check for design structural elements that may cause vulnerability to side-channel attacks: 
\textit{[If without else, Inferred Latches]}. If the code has these, we check whether the conditional statement contains any previously identified assets. The presence of these in a conditional statement without an \texttt{else} can result in information leaking. 

\subsubsection{Assertion-based}
We develop a custom template for System Verilog Assertions (SVAs) for selected CWEs and populate the template with information from AssetID. 
The formal tool verifies these SVAs for the RTL code, and \textit{falsified} assertions are sent to the next step, i.e., contextualization. We use VC Formal Property Verification (FPV)~\cite{vc_formal_vc_2024} for this.

For CWE 1231, we check for a signal containing lock bits that are modifiable when they should not be. 
From AssetID, we obtain the lock signal and the conditions under which it should be modified (the negation captures conditions when it should not be modifiable). 
For each lock signal and the corresponding conditions, we form the following template and populate it with the appropriate information:

\begin{lstlisting}[language=Verilog, numbers=none, frame=none ]
@([CLK_SENSE] [CLK])
[CONDITIONS_FOR_STABLE_LOCK]| => $stable([LOCK_SIGNAL]); 
\end{lstlisting}

For CWE 1233, we check for a security-sensitive signal that is missing lock bit protection. 
From AssetID, we obtain the security-sensitive register that should be protected, the lock signal that should be protecting it, and the reset conditions under which this protection mechanism is not applicable. 
For each lock signal, we form and populate the following template: 

\begin{lstlisting}[language=Verilog, numbers=none, frame=none ]
@([CLK_SENSE] [CLK])
disable iff ([RESET_CONDITIONS]) [LOCK_SIGNAL] == '1 |=> $stable([SECURITY_SENSITIVE_REGISTER]);
\end{lstlisting}

\noindent \autoref{fig:motivating-example-cwe1233-sa-outputs} captures an example of this assertion formation. 

For CWE 1244, we check for a privilege level signal that is escalated under conditions it should not have been. 
From AssetID, we obtain the privilege level signal, the correct conditions under which it should be escalated, the reset conditions of the module, the higher privilege level, and the signal containing the previous privilege level. 
A negation of escalation conditions provides the conditions under which the privilege signal should not be escalated. 
For this privilege signal, we form and populate the following assertion template: 

\begin{lstlisting}[language=Verilog, numbers=none, frame=none ]
@([CLK_SENSE] [CLK]) 
disable iff ([RESET_CONDITIONS])
~[CONDITIONS_FOR_PRIVILEGE_ESCALATION] |=>
    ([PRIVILEGE_SIGNAL] != [HIGH_PRIVILEGE] ||
    [PRIVILEGE_SIGNAL] == [PREVIOUS_PRIVILEGE] );  
\end{lstlisting}

\subsection{Contextualization}

First, the LLM reasons whether the reported linting violations or falsified properties pose a security issue pertinent to the CWE under consideration. 
If it reasons that there is a security issue, the LLM is prompted to explain why. 
The explanation and static analysis violation are the final outputs of \sol{} provided to the RTL designer.
The components of the prompt to the LLM are illustrated below:

\begin{lstlisting}[language=Python, numbers=none, frame=lines ]
system_prompt = <same as Assets Identificaiton>
user_prompt = "Consider the following Verilog code: <RTL source code> For each of the <static analysis outputs>, determine whether the <output> poses a security issue pertaining to CWE-<x> and provide an explanation if that is the case. If the violation does not pose a security issue, no explanation is needed. Here is the output <output from Static Analysis>. "
\end{lstlisting}

For experiments where we adopt prompt strategy \textit{v2} (details in~\autoref{subsec:pe-variations}), we use the response of the LLM from the first contextualization request and re-prompt the LLM to reason through each of its suggested security issues,  categorizing the violations as \textit{insecure} only if very confident.
\autoref{fig:motivating-example-cwe1191-contextualization} and~\autoref{fig:motivating-example-cwe1233-contextualization} show examples of contextualization. 

\section{Experimental Details\label{sec:experiments}}


\subsection{Dataset}
\label{subsec:dataset}

Our dataset consists of 4 open-source RISC-V based SoCs: Hack@DAC 2021's OpenPiton buggy SoC (H@DAC-21)~\cite{noauthor_hack-eventhackatdac21_2024}, OpenTitan~\cite{lowrisc_contributors_open_2023}, Hummingbirdv2 E203 (E203)~\cite{nuclei_system_technology_hummingbirdv2_2022} and VeerWolf~\cite{chipsalliance_veerwolf_2024}. Their details are mentioned in~\autoref{tbl:dataset}.

\begin{table}[h]

\footnotesize
\centering
\setlength{\tabcolsep}{2.5pt}

\caption{Dataset of open-source SoCs scanned for relevant CWEs.}
\label{tbl:dataset}

\begin{tabular}{lL{4.65cm}rr}

\toprule
SoC & Description & \#Mods & \#LoCs \\

\midrule
\midrule

H@DAC-21~\cite{noauthor_hack-eventhackatdac21_2024}    & OpenPiton SoC (CVA6 core) for Hack@DAC 2021 competition & 63      & 15k \\
\midrule

OpenTitan~\cite{lowrisc_contributors_open_2023}    & Silicon Root of Trust project (Ibex core) & 359     & 171k \\
\midrule

E203~\cite{nuclei_system_technology_hummingbirdv2_2022}         & Hummingbirdv2 E203 core and SoC & 76      & 27k \\
\midrule

VeerWolf~\cite{chipsalliance_veerwolf_2024}     &  FuseSoC-based platform for VeeR cores & 34      & 11k\\

\bottomrule
 
\end{tabular}
\end{table}


\subsection{Hardware Common Weakness Enumerations (CWEs)}
\label{subsec:cwes}

Security-related issues that arise because of hardware bugs are taxonomized as Common Weakness Enumerations (CWEs). MITRE~\cite{the_mitre_corporation_mitre_cwe_2022} works with academia and industry to develop a list of CWEs that represent categories of vulnerabilities.
A weakness is an element in a digital product’s software, firmware, hardware, or service that can be exploited for malicious purposes. We develop \sol{} for 5 CWEs selected from the list of Most Important Hardware CWEs published by MITRE. We selected the ones that had coded examples on MITRE's website and were present in the H@DAC-21 SoC to have some ground-truth for validation of initial prototyping. The 5 CWEs covered are described in~\autoref{tbl:cwes}.

\begin{table}[t]

\footnotesize
\centering
\setlength{\tabcolsep}{2pt}

\caption{CWEs covered by \sol{}. Selected from MITRE's list of Most Important Hardware CWEs.}
\label{tbl:cwes}

\begin{tabular}{ll}

\toprule
CWE & Description\\

\midrule
\midrule

1191& On-Chip Debug and Test Interface With Improper Access Control \\
\midrule
1231& Improper Prevention of Lock Bit Modification \\
\midrule
1233& Security-Sensitive Hardware Controls with Missing Lock Protection \\
\midrule
1244& Internal Asset Exposed to Unsafe Debug Access Level or State \\
\midrule
1300& Improper Protection of Physical Side Channels \\

\bottomrule
 
\end{tabular}
\end{table}

\subsection{Prompt Variations}
\label{subsec:pe-variations}
The performance of LLMs is dependent on the quality of prompts and examples of the correct solutions to the task at hand.
To study the extent to which in-context learning~\cite{zhou_large_2022} and insistence on reasoning helps in \sol{}'s performance, we guide the LLM through 4 prompt variations.
The 4 variations are formed using combinations of 2 improvements.
The \textit{first improvement} helps in the Assets Identification and Contextualization phases by providing a comprehensive example of a hardware security bug that captures the CWE.
The \textit{second improvement} helps in the Contextualization phase by asking the LLM to think again about its initial assessments of whether the reported violation poses a security risk.

\subsubsection{Variation \textit{v0} (baseline)} is the zero-shot implementation for \sol{}. It contains no in-context learning or request to re-evaluate outputs and forms the baseline to compare improvements in performance with variations \textit{v1}, \textit{v2} and \textit{v3}.

\subsubsection{Variation \textit{v1}} uses the \textit{first improvement} only. The examples and descriptions of bugs for each CWE are taken from the MITRE's website. Each example appears in the prompt after the LLM is given its role and information about the CWE it is going to look for.
It consists of the bug in RTL form, the explanation of the security issues because of the bug and the relevant security assets. Here is an example for the prompt variation \textit{v1} appended to the baseline \textit{v0} for CWE 1231 (full example in Appendix~\autoref{subsec:prompt-variation-appendix}):


\noindent\rule{\linewidth}{0.4pt} \\
\color{blue}
Here is an example of CWE-1231 with code from the register locks module:\\
\color{olive}
always @(posedge clk\_i) begin\\
\indent if(~(rst\_ni \&\& ~jtag\_unlock \&\& ~rst\_9)) begin\\
\indent\indent for (j=0; j < 6; j=j+1) begin\\
\indent\indent\indent reglk\_mem[j] <= 'h0;\\
\color{magenta}
\indent \indent \indent \indent \indent \indent <Explanation of the security issue>\\
\color{blue}
In this example the lock signal is reglk\_mem and the correct conditions for changing lock signals are (rst\_ni \&\& ~jtag\_unlock).\\
\color{black}
\noindent\rule{\linewidth}{0.4pt}


\subsubsection{Variation \textit{v2}} uses the \textit{second improvement} only. The model is ``given time to think'' so that it can double check its initial assessment. It is asked to use inner monologue to go over its reasoning process.
It looks the same for all CWEs and appears in the user prompt for Contextualization. First, the LLM is asked to go over the violations and assess which of them are actually insecure and which are not. This output is then sent back to the LLM to simulate a chat. An example of the instruction to `reason' for CWE-1231 is shown below:


\noindent\rule{\linewidth}{0.4pt} \\
\color{blue}
Go over the previously provided response and reason about the provided explanation for each falsified property. Only categorize the falsified property as insecure if you are confident in your assessment. Here is the `falsified\_properties' object:\\
\color{magenta}
\indent \indent \indent \indent \indent <string of falsified assertions information>\\
\color{black}
\noindent\rule{\linewidth}{0.4pt}



\subsubsection{Variation \textit{v3}} uses both, \textit{first} and \textit{second improvements}.

\subsection{Large Language Models (LLMs)}
We use 2 OpenAI LLMs, \textit{gpt-4o-mini-2024-07-18} and \textit{gpt-4o-2024-08-06}~\cite{openai_gpt-4o_2024}, to conduct out experiments and evaluate if using a more powerful LLM makes a difference for our setup.
\textit{gpt-4o-mini} is OpenAI's most advanced model in the small models category.
\textit{gpt-4o} is OpenAI's most advanced GPT model and is slower and more advanced than \textit{gpt-4o}.
Both these models have the ability to give structured outputs consistent with the object structure provided.


\section{Results\label{sec:results}}

We evaluated \sol{} on the 4 SoCs for 5 CWEs, 4 prompt variations and 2 LLMs. The results are summarized in~\autoref{tbl:results-summary}. In total, 545 instances are flagged across the 160 experiments out of which 51\% are potential CWEs. 2026 assertions were formed and 12,554 assets were checked in the process. Since we do not
know the real number of bugs (we can only confirm or deny a specific bug's presence after flagging), 
it is not possible to calculate a proper Recall or Accuracy score.
Therefore, we rely on Precision i.e. (\# true positives / \# predicted positive (flagged)) as the metric to evaluate performance.
For our recommended combination of using \textit{gpt-4o} with \textit{v3}, 35 of the 40 flagged instances are plausible CWEs, providing a precision of 87.5\%.
On average, for a given SoC being searched for a particular CWE, choosing the appropriate prompt variation, there are 3.4 violations reported. We evaluated the violations manually, with our author-confirmed violations providing the True Positives count in~\autoref{tbl:results-summary}.

There is significant variation in \sol{}'s success based on the CWE. It performs the best on CWEs 1191, 1244 and 1300 with precisions of 1, 0.8 and 0.91 and worse on CWEs 1231 and 1233 with precisions of 0.28 and 0.45.
CWE 1231 was harder to identify because of poor asset identification. It was difficult for the LLM to identify the lock bit signal that should remain stable and instead kept identifying control and status registers.
CWE 1233 was harder to accurately identify because of the range of possibilities of how a security sensitive signal may be protected in RTL.
It may be protected by an if condition or by being `anded' (\&) with another signal or by being assigned a signal which may have some protection.

\begin{table}[h]
\footnotesize

\centering
\setlength{\tabcolsep}{4pt}

\caption{Results Summary.
Precision = True Positives (TPs) / Flagged,
False Discovery Rate (FDR) = False Positives / Flagged, 
Assets = number of assets identified ,
Assertions = number of assertions formed from custom templates for each CWE. The best results are emboldened.}
\label{tbl:results-summary}

\begin{tabular}{ccrrrrrr}

\toprule
\multicolumn{1}{l}{CWE} & Variation & \multicolumn{1}{l}{Flagged} & \multicolumn{1}{l}{TPs} & \multicolumn{1}{l}{Precision} & \multicolumn{1}{l}{FDR} & \multicolumn{1}{l}{Assets} & \multicolumn{1}{l}{Assertions} \\

\midrule
\midrule

\textbf{\multirow{4}{*}{1191} }& v0 & 12	&	12	&	1.00	&	0.00	&	511	&	- \\
 & v1 & 11	&	11	&	1.00	&	0.00	&	580	&	- \\
 & \textbf{v2} & 13	&	\textbf{13}	&	\textbf{1.00}	&	0.00	&	511	&	- \\
 & v3 & 9	&	9	&	1.00	&	0.00	&	580	&	- \\
\cmidrule{3-8}
 & & 45	&	45	&	1.00	&	0.00	&	2182	&	- \\
\midrule

\multirow{4}{*}{1231} &	v0	&	15	&	3	&	0.20	&	0.80	&	33	&	33	\\
&	\textbf{v1}	&	8	&	\textbf{6}	&	\textbf{0.75}	&	0.25	&	11	&	11	\\
&	v2	&	24	&	2	&	0.08	&	0.92	&	33	&	33	\\
&	v3	&	6	&	4	&	0.67	&	0.33	&	11	&	11	\\
 \cmidrule{3-8}
&		&	53	&	15	&	0.28	&	0.72	&	88	&	88	\\
\midrule
 
\multirow{4}{*}{1233} &	v0	&	143	&	\textbf{55}	&	0.38	&	0.62	&	669	&	587	\\
&	v1	&	114	&	48	&	0.42	&	0.58	&	406	&	355	\\
&	v2	&	68	&	35	&	0.51	&	0.49	&	669	&	587	\\
&	\textbf{v3}	&	89	&	50	&	\textbf{0.56}	&	0.44	&	406	&	355	\\
 \cmidrule{3-8}
&		&	414	&	188	&	0.45	&	0.55	&	2150	&	1884	\\
\midrule
 
\multirow{4}{*}{1244} &	v0	&	5	&	3	&	0.60	&	0.40	&	24	&	12	\\
&	\textbf{v1}	&	3	&	\textbf{3}	&	\textbf{1.00}	&	0.00	&	22	&	15	\\
&	v2	&	1	&	1	&	1.00	&	0.00	&	24	&	12	\\
&	v3	&	1	&	1	&	1.00	&	0.00	&	22	&	15	\\
 \cmidrule{3-8}
&		&	10	&	8	&	0.80	&	0.20	&	92	&	54	\\
\midrule
 
\multirow{4}{*}{1300} &	v0	&	8	&	7	&	0.88	&	0.13	&	2053	&	-	\\
&	v1	&	9	&	8	&	0.89	&	0.11	&	1968	&	-	\\
&	v2	&	3	&	3	&	1.00	&	0.00	&	2053	&	-	\\
&	\textbf{v3}	&	3	&	\textbf{3}	&	\textbf{1.00}	&	0.00	&	1968	&	-	\\
 \cmidrule{3-8}
&		&	23	&	21	&	0.91	&	0.09	&	8042	&	-	\\

 \midrule
&		&	545	&	277	&	0.508	&	0.49	&	12554	&	2026 \\
\bottomrule
 
\end{tabular}
\end{table}

\subsection{Analysis}
\label{subsec:analysis}

\begin{figure*}[tbh]
    \centering
    \includegraphics[width=\linewidth]{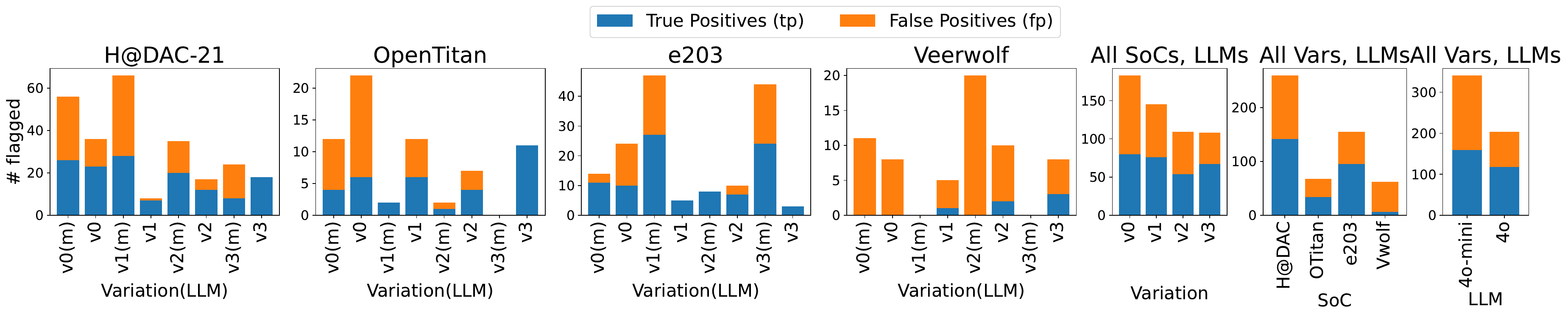}
    \caption{Classification of instances flagged by \sol{}. Stacked bar shows the number of instances identified by \sol{} that pose security issues as true positives (tp) and those that do not as false positives (fp). The numbers are summed up for all CWEs.}
    \vspace{-2mm}
    \label{fig:flagged-analysis}
\end{figure*}

\begin{figure*}[tbh]
    \centering
    \includegraphics[width=\linewidth]{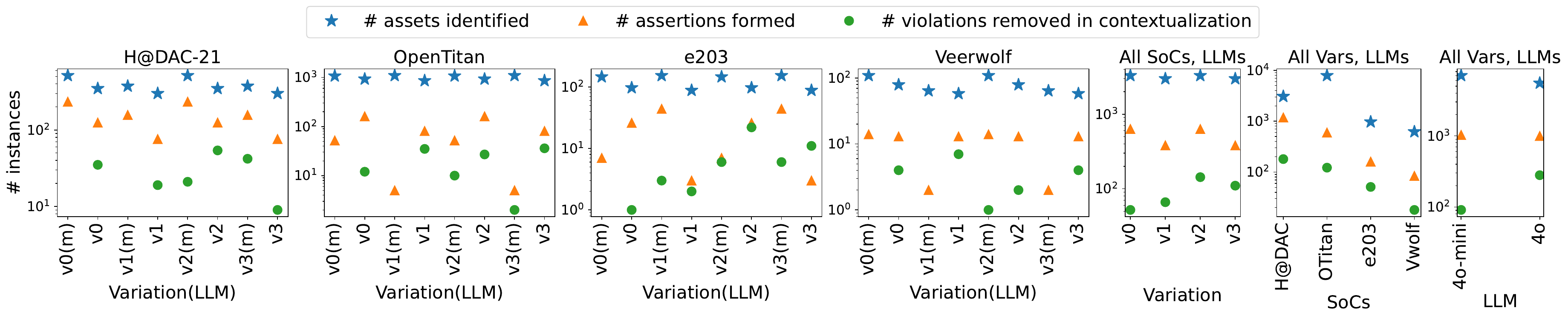}
    \caption{Intermediary outputs during \sol{}'s operation. \# assets identified is number of security relevant signals identified during Assets Identification, \# assertions formed is the number of assertions formed during Static Analysis. \# violations removed in contextualization is the number of violations that the LLM reasoned as not posing a security threat. The numbers are summed up for all CWEs.}
    \label{fig:other-stats-analysis}
    \vspace{-2mm}
\end{figure*}

\subsubsection{Impact of Prompt Variations}
Prompt variations were successful in improving performance. As shown in~\autoref{fig:flagged-analysis}-(Variation), variation \textit{v1} which introduces an example of CWE to guide the LLM, results in finding less false positives. The number of true positives decreases slightly from 80 to 76 while the precision improves from 0.44 to 0.52 in comparison to the baseline \textit{v0}.
Introducing \textit{v2}, which asks the LLM to reason about its initial assessment during contextualization, shows improvement as well but to a lesser extent. The number of true positives decreases from 80 to 54 but the precision increases from 0.44 to 0.50 in comparison to the baseline \textit{v0}. \textit{v2} significantly reduced false positives from 103 to 55.
The best performing variation is \textit{v3} with highest precision of 0.62, a 42\% improvement over the baseline.
As \textit{v3} is a combination of \textit{v1} and \textit{v2}, it seems that \textit{v1} can improve the `search' of \sol{} by identifying assets that are more relevant to the CWE and then \textit{v2} prunes out the remaining false positives. This can be seen in~\autoref{fig:other-stats-analysis}-(Variation) with an increase in removal of violations in contextualization for \textit{v3}.

\subsubsection{Which SoCs were better analyzed?}
\sol{}'s performance significantly varies with the SoC as illustrated in~\autoref{fig:flagged-analysis}-(SoC). It performed the best on e203 with a precision of 0.61 and the worst on Veerwolf with a precision of 0.10. The main culprit for poor performance on Veerwolf is the large number of false positives for CWE-1231. \sol{} kept misidentifying a lot of the control and status registers as lock registers. Hack@DAC-21 performs the second best with a precision of 0.55, which is expected because the guiding examples we use in \textit{v1} are inspired from the Hack@DAC-21 SoC. OpenTitan has a precision close to that of Hack@DAC-21, 0.50, which validates our approach -- OpenTitan is significantly larger than Hack@DAC-21 in size and complexity.

Another area \sol{} had performance issues in with Veerwolf and OpenTitan came with the forming of meaningful assertions. The ratio of assertion formation to number of assets identified for Hack@DAC-21 is significantly larger, as shown in~\autoref{fig:other-stats-analysis}-(SoC). This is perhaps due to the guiding examples being based on Hack@DAC-21.

\subsubsection{Which LLM performed better?}
\sol{} performed better with \textit{gpt-4o} (precision 0.58) compared to \textit{gpt-40-mini} (precision 0.47).
\textit{gpt-4o} flags less violations 204 vs. 341, less true positives 118 vs. 159, but also, less false positives 86 vs. 182, than \textit{gpt-4o-mini}. 
This difference is highlighted when considering \textit{v3} only: \textit{gpt-4o} detects more true positives 35 vs. 32 and has a higher precision 0.88 vs. 0.47.


\section{Discussion\label{sec:dis}}

Our work shows that a combination of LLMs and static analysis can provide the `best of both worlds', with each tool helping to overcome the limitations of the other.
Detecting vulnerabilities with only LLMs leads to a large number of false positives and limited confidence on the outputs because there is no verification.
Conversely, detecting vulnerabilities with formal tools requires a lot of expertise and precise information which can be hard to obtain.
We demonstrate that the limited confidence on the outputs of LLMs can be improved by verification through static analysis tools and the requirement of specific information can be provided by the LLM to some extent. The precision of 0.51 for \sol{} (for all experiments combined) can be interpreted as a 51\% confidence in the identification of a security issue. 

Linting checks and assertions are very different in their nature. Lint rules are harder to map to security issues and assertions are harder to form correctly. Which of these work better in our flow is still an open question. The lint checks that appeared often in flagged instances were related to improper range indexes, concatenations and `if' statements missing `else' statements.
While the confidence of reported violations is high, there are more false positives than would be ideal. The main culprit is liberal asset identification by the LLM. Providing the CWE information is not enough to constrain the number of assets the LLM identifies. Providing information regarding the operation of the specific modules and the security objectives could improve results.
Another reason for false positives is the difficulty in forming correct assertions. Incorrect conditions identified by the LLM lead to failing assertions which flag violations erroneously.
Another limitation lies in the manual evaluation of flagged violations. We assessed all violations through visual inspection which has a possibility of being incorrect. Each violation, however, is accompanied by a failing assertion or violation which gives credence to the classification.


\section{Related Prior Works}

Few works have explored using both LLMs and Static Analysis components to identify RTL security bugs.
SoCureLLM~\cite{tarek_socurellm_2024} is an LLM-driven approach for large-scale System-on-Chip security verification and policy generation. While they use LLMs for security policy generation and security policy violation, they do not use static tools for security violation in their solution.
Flag-RTL~\cite{ahmad_flag_2023} uses LLMs for bug detection in RTL but uses static analysis differently. A front end parser localizes the LLM's search to particular parts of the code, but no static analysis tool is used for verification.
Self-HWDebug~\cite{akyash_self-hwdebug_2024} automates LLM self-instructing for hardware security verification. This work uses known bugs and their CWEs to generate instructions for debugging and repair and there is no use of static analysis. While these works are related to \sol{}, there are significant differences which do not allow for a direct comparison. None of the works use LLMs and Static Analysis together in equal levels of significance and none explore their tools on unknown bugs.

\section{Conclusion\label{sec:conclusions}}

This work combines LLMs and Static Analysis for hardware security bug detection. \sol{}, by using LLMs equipped with in-context learning and requests for `thinking again', has reasonable success in finding unknown bugs with an empirical precision of 0.88. On average, \sol{} reported 3.4 violations per CWE for a given SoC, out of which 1.7 were evaluated to be plausible security issues.  This  is demonstrated over 5 of the most important Hardware CWEs, with the best performance on CWE 1191 (100\% precision) and worst on CWE 1231 (28\% precision). We catered the static analysis approach to the nature of the CWEs to show 2 techniques that do similarly well i.e., linting and assertions.
Future work could investigate the application of reasoning models such as OpenAI o1  not available during our experimentation, as well as examine support for more CWEs. 


\section{Acknowledgments\label{sec:acks}}
This research work is supported in part by a gift from Intel Corporation. This work does not in any way constitute an Intel endorsement of a product or supplier.
We thank Verific Design Automation for generously providing academic access to linkable libraries, examples, and documentation for their RTL parsers.


\bibliographystyle{IEEEtran}
\bibliography{Lit/benhamram}

\begin{thebibliography}{10}
\providecommand{\url}[1]{#1}
\csname url@samestyle\endcsname
\providecommand{\newblock}{\relax}
\providecommand{\bibinfo}[2]{#2}
\providecommand{\BIBentrySTDinterwordspacing}{\spaceskip=0pt\relax}
\providecommand{\BIBentryALTinterwordstretchfactor}{4}
\providecommand{\BIBentryALTinterwordspacing}{\spaceskip=\fontdimen2\font plus
\BIBentryALTinterwordstretchfactor\fontdimen3\font minus \fontdimen4\font\relax}
\providecommand{\BIBforeignlanguage}[2]{{%
\expandafter\ifx\csname l@#1\endcsname\relax
\typeout{** WARNING: IEEEtran.bst: No hyphenation pattern has been}%
\typeout{** loaded for the language `#1'. Using the pattern for}%
\typeout{** the default language instead.}%
\else
\language=\csname l@#1\endcsname
\fi
#2}}
\providecommand{\BIBdecl}{\relax}
\BIBdecl

\bibitem{dessouky_hardfails_2019}
\BIBentryALTinterwordspacing
G.~Dessouky, D.~Gens, P.~Haney, G.~Persyn, A.~Kanuparthi, H.~Khattri, J.~M. Fung, A.-R. Sadeghi, and J.~Rajendran, ``\BIBforeignlanguage{en}{{HardFails}: {Insights} into {Software}-{Exploitable} {Hardware} {Bugs}},'' 2019, pp. 213--230. [Online]. Available: \url{https://www.usenix.org/conference/usenixsecurity19/presentation/dessouky}
\BIBentrySTDinterwordspacing

\bibitem{mitra_post-silicon_2010}
\BIBentryALTinterwordspacing
S.~Mitra, S.~A. Seshia, and N.~Nicolici, ``Post-silicon validation opportunities, challenges and recent advances,'' in \emph{Proceedings of the 47th {Design} {Automation} {Conference}}, ser. {DAC} '10.\hskip 1em plus 0.5em minus 0.4em\relax New York, NY, USA: Association for Computing Machinery, Jun. 2010, pp. 12--17. [Online]. Available: \url{https://dl.acm.org/doi/10.1145/1837274.1837280}
\BIBentrySTDinterwordspacing

\bibitem{ray_invited_2019}
S.~Ray, N.~Ghosh, R.~Masti, A.~Kanuparthi, and J.~Fung, ``{INVITED}: {Formal} {Verification} of {Security} {Critical} {Hardware}-{Firmware} {Interactions} in {Commercial} {SoCs},'' in \emph{2019 56th {ACM}/{IEEE} {Design} {Automation} {Conference} ({DAC})}, Jun. 2019, pp. 1--4, iSSN: 0738-100X.

\bibitem{he_soc_2019}
\BIBentryALTinterwordspacing
J.~He, X.~Guo, T.~Meade, R.~Dutta, Y.~Zhao, and Y.~Jin, ``\BIBforeignlanguage{en}{{SoC} interconnection protection through formal verification},'' \emph{\BIBforeignlanguage{en}{Integration}}, vol.~64, pp. 143--151, Jan. 2019. [Online]. Available: \url{https://www.sciencedirect.com/science/article/pii/S016792601830289X}
\BIBentrySTDinterwordspacing

\bibitem{hur_difuzzrtl_2021}
J.~Hur, S.~Song, D.~Kwon, E.~Baek, J.~Kim, and B.~Lee, ``{DifuzzRTL}: {Differential} {Fuzz} {Testing} to {Find} {CPU} {Bugs},'' in \emph{2021 {IEEE} {Symposium} on {Security} and {Privacy} ({SP})}, May 2021, pp. 1286--1303, iSSN: 2375-1207.

\bibitem{trippel_fuzzing_2022}
\BIBentryALTinterwordspacing
T.~Trippel, K.~G. Shin, A.~Chernyakhovsky, G.~Kelly, D.~Rizzo, and M.~Hicks, ``\BIBforeignlanguage{en}{Fuzzing {Hardware} {Like} {Software}},'' 2022, pp. 3237--3254. [Online]. Available: \url{https://www.usenix.org/conference/usenixsecurity22/presentation/trippel}
\BIBentrySTDinterwordspacing

\bibitem{ardeshiricham_register_2017}
A.~Ardeshiricham, W.~Hu, J.~Marxen, and R.~Kastner, ``Register transfer level information flow tracking for provably secure hardware design,'' in \emph{Design, {Automation} {Test} in {Europe} {Conference} {Exhibition} ({DATE}), 2017}, Mar. 2017, pp. 1691--1696, iSSN: 1558-1101.

\bibitem{hu_hardware_2021}
\BIBentryALTinterwordspacing
W.~Hu, A.~Ardeshiricham, and R.~Kastner, ``Hardware {Information} {Flow} {Tracking},'' \emph{ACM Computing Surveys}, vol.~54, no.~4, pp. 83:1--83:39, May 2021. [Online]. Available: \url{https://dl.acm.org/doi/10.1145/3447867}
\BIBentrySTDinterwordspacing

\bibitem{ahmad_dont_2022}
\BIBentryALTinterwordspacing
B.~Ahmad, W.-K. Liu, L.~Collini, H.~Pearce, J.~M. Fung, J.~Valamehr, M.~Bidmeshki, P.~Sapiecha, S.~Brown, K.~Chakrabarty, R.~Karri, and B.~Tan, ``Don't {CWEAT} {It}: {Toward} {CWE} {Analysis} {Techniques} in {Early} {Stages} of {Hardware} {Design},'' in \emph{Proceedings of the 41st {IEEE}/{ACM} {International} {Conference} on {Computer}-{Aided} {Design}}, ser. {ICCAD} '22.\hskip 1em plus 0.5em minus 0.4em\relax New York, NY, USA: Association for Computing Machinery, Dec. 2022, pp. 1--9. [Online]. Available: \url{https://doi.org/10.1145/3508352.3549369}
\BIBentrySTDinterwordspacing

\bibitem{bidmeshki_hunting_2021}
M.~M. Bidmeshki, Y.~Zhang, M.~Zaman, L.~Zhou, and Y.~Makris, ``Hunting {Security} {Bugs} in {SoC} {Designs}: {Lessons} {Learned},'' \emph{IEEE Design \& Test}, vol.~38, no.~1, pp. 22--29, Feb. 2021.

\bibitem{ahmad_flag_2023}
\BIBentryALTinterwordspacing
B.~Ahmad, B.~Tan, R.~Karri, and H.~Pearce, ``{FLAG}: {Finding} {Line} {Anomalies} (in code) with {Generative} {AI},'' Jun. 2023, arXiv:2306.12643 [cs]. [Online]. Available: \url{http://arxiv.org/abs/2306.12643}
\BIBentrySTDinterwordspacing

\bibitem{tarek_socurellm_2024}
\BIBentryALTinterwordspacing
S.~Tarek, D.~Saha, S.~K. Saha, M.~Tehranipoor, and F.~Farahmandi, ``{SoCureLLM}: {An} {LLM}-driven {Approach} for {Large}-{Scale} {System}-on-{Chip} {Security} {Verification} and {Policy} {Generation},'' 2024, publication info: Preprint. [Online]. Available: \url{https://eprint.iacr.org/2024/983}
\BIBentrySTDinterwordspacing

\bibitem{akyash_self-hwdebug_2024}
\BIBentryALTinterwordspacing
M.~Akyash and H.~M. Kamali, ``Self-{HWDebug}: {Automation} of {LLM} {Self}-{Instructing} for {Hardware} {Security} {Verification},'' May 2024, arXiv:2405.12347. [Online]. Available: \url{http://arxiv.org/abs/2405.12347}
\BIBentrySTDinterwordspacing

\bibitem{vc_spyglass_lint_synopsys_2022}
\BIBentryALTinterwordspacing
V.~S. Lint, ``\BIBforeignlanguage{en}{Synopsys {VC} {SpyGlass} {Lint}},'' 2022. [Online]. Available: \url{https://www.synopsys.com/verification/static-and-formal-verification/vc-spyglass/vc-spyglass-lint.html}
\BIBentrySTDinterwordspacing

\bibitem{jasperlint_jasper_2022}
\BIBentryALTinterwordspacing
jasperlint, ``\BIBforeignlanguage{en}{Jasper {Superlint} {App}},'' 2022. [Online]. Available: \url{https://www.cadence.com/en_US/home/tools/system-design-and-verification/formal-and-static-verification/jasper-gold-verification-platform/jaspergold-superlint-app.html}
\BIBentrySTDinterwordspacing

\bibitem{noauthor_vc_2022}
\BIBentryALTinterwordspacing
``\BIBforeignlanguage{en}{{VC} {Formal}},'' 2022. [Online]. Available: \url{https://www.synopsys.com/verification/static-and-formal-verification/vc-formal.html}
\BIBentrySTDinterwordspacing

\bibitem{cadence_jasper_2022}
\BIBentryALTinterwordspacing
Cadence, ``\BIBforeignlanguage{en}{Jasper {RTL} {Apps} {\textbar} {Cadence}},'' Jul. 2022. [Online]. Available: \url{https://www.cadence.com/en_US/home/tools/system-design-and-verification/formal-and-static-verification/jasper-gold-verification-platform.html}
\BIBentrySTDinterwordspacing

\bibitem{hansson_continuous_2014}
D.~Hansson, ``Continuous {Linting} with {Automatic} {Debug},'' in \emph{2014 15th {International} {Microprocessor} {Test} and {Verification} {Workshop}}, Dec. 2014, pp. 70--72, iSSN: 2332-5674.

\bibitem{mcnutt_linting_2018}
A.~McNutt and G.~Kindlmann, ``\BIBforeignlanguage{en}{Linting for {Visualization}: {Towards} a {Practical} {Automated} {Visualization} {Guidance} {System}},'' 2018.

\bibitem{woodcock_formal_2009}
\BIBentryALTinterwordspacing
J.~Woodcock, P.~G. Larsen, J.~Bicarregui, and J.~Fitzgerald, ``Formal methods: {Practice} and experience,'' \emph{ACM Comput. Surv.}, vol.~41, no.~4, pp. 19:1--19:36, Oct. 2009. [Online]. Available: \url{https://dl.acm.org/doi/10.1145/1592434.1592436}
\BIBentrySTDinterwordspacing

\bibitem{chen_evaluating_2021}
\BIBentryALTinterwordspacing
M.~Chen, J.~Tworek, H.~Jun, Q.~Yuan, H.~P. d.~O. Pinto, J.~Kaplan, H.~Edwards, Y.~Burda, N.~Joseph, G.~Brockman, A.~Ray, R.~Puri, G.~Krueger, M.~Petrov, H.~Khlaaf, G.~Sastry, P.~Mishkin, B.~Chan, S.~Gray, N.~Ryder, M.~Pavlov, A.~Power, L.~Kaiser, M.~Bavarian, C.~Winter, P.~Tillet, F.~P. Such, D.~Cummings, M.~Plappert, F.~Chantzis, E.~Barnes, A.~Herbert-Voss, W.~H. Guss, A.~Nichol, A.~Paino, N.~Tezak, J.~Tang, I.~Babuschkin, S.~Balaji, S.~Jain, W.~Saunders, C.~Hesse, A.~N. Carr, J.~Leike, J.~Achiam, V.~Misra, E.~Morikawa, A.~Radford, M.~Knight, M.~Brundage, M.~Murati, K.~Mayer, P.~Welinder, B.~McGrew, D.~Amodei, S.~McCandlish, I.~Sutskever, and W.~Zaremba, ``Evaluating {Large} {Language} {Models} {Trained} on {Code},'' Jul. 2021, arXiv:2107.03374 [cs]. [Online]. Available: \url{http://arxiv.org/abs/2107.03374}
\BIBentrySTDinterwordspacing

\bibitem{thakur_verigen_2024}
\BIBentryALTinterwordspacing
S.~Thakur, B.~Ahmad, H.~Pearce, B.~Tan, B.~Dolan-Gavitt, R.~Karri, and S.~Garg, ``{VeriGen}: {A} {Large} {Language} {Model} for {Verilog} {Code} {Generation},'' \emph{ACM Trans. Des. Autom. Electron. Syst.}, vol.~29, no.~3, pp. 46:1--46:31, Apr. 2024. [Online]. Available: \url{https://doi.org/10.1145/3643681}
\BIBentrySTDinterwordspacing

\bibitem{ahmad_hardware_2024}
\BIBentryALTinterwordspacing
B.~Ahmad, S.~Thakur, B.~Tan, R.~Karri, and H.~Pearce, ``On {Hardware} {Security} {Bug} {Code} {Fixes} by {Prompting} {Large} {Language} {Models},'' \emph{IEEE Transactions on Information Forensics and Security}, vol.~19, pp. 4043--4057, 2024. [Online]. Available: \url{https://ieeexplore.ieee.org/abstract/document/10462177}
\BIBentrySTDinterwordspacing

\bibitem{li_llm-assisted_2024}
\BIBentryALTinterwordspacing
Z.~Li, S.~Dutta, and M.~Naik, ``{LLM}-{Assisted} {Static} {Analysis} for {Detecting} {Security} {Vulnerabilities},'' Nov. 2024, arXiv:2405.17238. [Online]. Available: \url{http://arxiv.org/abs/2405.17238}
\BIBentrySTDinterwordspacing

\bibitem{chapman_interleaving_2024}
\BIBentryALTinterwordspacing
P.~J. Chapman, C.~Rubio-González, and A.~V. Thakur, ``\BIBforeignlanguage{en}{Interleaving {Static} {Analysis} and {LLM} {Prompting}},'' in \emph{\BIBforeignlanguage{en}{Proceedings of the 13th {ACM} {SIGPLAN} {International} {Workshop} on the {State} {Of} the {Art} in {Program} {Analysis}}}.\hskip 1em plus 0.5em minus 0.4em\relax Copenhagen Denmark: ACM, Jun. 2024, pp. 9--17. [Online]. Available: \url{https://dl.acm.org/doi/10.1145/3652588.3663317}
\BIBentrySTDinterwordspacing

\bibitem{li_enhancing_2024}
\BIBentryALTinterwordspacing
H.~Li, Y.~Hao, Y.~Zhai, and Z.~Qian, ``Enhancing {Static} {Analysis} for {Practical} {Bug} {Detection}: {An} {LLM}-{Integrated} {Approach},'' \emph{Enhancing Static Analysis for Practical Bug Detection: An LLM-Integrated Approach (Artifact)}, vol.~8, no. OOPSLA1, pp. 111:474--111:499, Apr. 2024. [Online]. Available: \url{https://dl.acm.org/doi/10.1145/3649828}
\BIBentrySTDinterwordspacing

\bibitem{the_mitre_corporation_mitre_cwe_2022}
\BIBentryALTinterwordspacing
T.~M.~C. (MITRE), ``{CWE} - {CWE} {Most} {Important} {Hardware} {Weaknesses},'' 2022. [Online]. Available: \url{https://cwe.mitre.org/scoring/lists/2021_CWE_MIHW.html}
\BIBentrySTDinterwordspacing

\bibitem{verific_verific_2022}
\BIBentryALTinterwordspacing
Verific, ``\BIBforeignlanguage{en-US}{Verific {Design} {Automation}},'' 2022. [Online]. Available: \url{https://www.verific.com/}
\BIBentrySTDinterwordspacing

\bibitem{vc_formal_vc_2024}
\BIBentryALTinterwordspacing
V.~Formal, ``\BIBforeignlanguage{en}{{VC} {Formal}: {Formal} {Verification} {Solution} {\textbar} {Synopsys}},'' 2024. [Online]. Available: \url{https://www.synopsys.com/verification/static-and-formal-verification/vc-formal.html}
\BIBentrySTDinterwordspacing

\bibitem{noauthor_hack-eventhackatdac21_2024}
\BIBentryALTinterwordspacing
``{HACK}-{EVENT}/hackatdac21,'' Apr. 2024, original-date: 2023-07-15T20:58:02Z. [Online]. Available: \url{https://github.com/HACK-EVENT/hackatdac21}
\BIBentrySTDinterwordspacing

\bibitem{lowrisc_contributors_open_2023}
\BIBentryALTinterwordspacing
lowRISC contributors, ``\BIBforeignlanguage{en}{Open source silicon root of trust ({RoT}) {\textbar} {OpenTitan}},'' 2023. [Online]. Available: \url{https://opentitan.org/}
\BIBentrySTDinterwordspacing

\bibitem{nuclei_system_technology_hummingbirdv2_2022}
\BIBentryALTinterwordspacing
N.~S. Technology, ``Hummingbirdv2 {E203} {Core} and {SoC} - {GitHub},'' May 2022, original-date: 2020-07-29T06:28:49Z. [Online]. Available: \url{https://github.com/riscv-mcu/e203_hbirdv2}
\BIBentrySTDinterwordspacing

\bibitem{chipsalliance_veerwolf_2024}
\BIBentryALTinterwordspacing
chipsalliance, ``{VeeRwolf},'' Nov. 2024, original-date: 2019-08-07T15:24:36Z. [Online]. Available: \url{https://github.com/chipsalliance/VeeRwolf}
\BIBentrySTDinterwordspacing

\bibitem{zhou_large_2022}
\BIBentryALTinterwordspacing
Y.~Zhou, A.~I. Muresanu, Z.~Han, K.~Paster, S.~Pitis, H.~Chan, and J.~Ba, ``Large {Language} {Models} {Are} {Human}-{Level} {Prompt} {Engineers},'' Nov. 2022, arXiv:2211.01910 [cs]. [Online]. Available: \url{http://arxiv.org/abs/2211.01910}
\BIBentrySTDinterwordspacing

\bibitem{openai_gpt-4o_2024}
\BIBentryALTinterwordspacing
OpenAI, ``\BIBforeignlanguage{en-US}{{GPT}-4o},'' May 2024. [Online]. Available: \url{https://openai.com/index/hello-gpt-4o/}
\BIBentrySTDinterwordspacing

\end{thebibliography}


\appendix

\section{Appendix\label{sec:appendix}}

\subsection{Scalability and Cost}
On average, each experiment took 163 seconds to run. This time includes the complete flow of running the LLM and static analysis tools from the identification of relevant RTL to the outputs of location and explanation of bugs.
In total, 160 experiments were run in 7.26 hours. There is a relation between the lines of code and the time taken. The dependency is linear in the log-log scale and produces follows the relation $time \propto loc^{0.53}$ approximately. This shows that the time taken grows proportional to the square root of the amount of code being analyzed, indicating a scalable approach. 
On average, each experiment cost \$0.055 while using gpt-4o-mini and \$0.379 while using gpt-4o. In total all experiments cost \$35 to run, catering to 40.2M input tokens.

\subsection{Propmpt variation \textit{v1}\label{subsec:prompt-variation-appendix}}

The examples and descriptions of bugs for each CWE are taken from the MITRE's website. Each example appears in the prompt after the LLM is given its role and information about the CWE it is going to look for.
It consists of the bug in RTL form, the explanation of the security issues because of the bug and the relevant security assets. Here is an example for the prompt variation \textit{v1} appended to the baseline \textit{v0} for CWE 1231.

\noindent\rule{\linewidth}{0.4pt} \\
\color{blue}
Here is an example of CWE-1231 with code from the register locks module:\\
\color{olive}
"""\\
always @(posedge clk\_i) begin\\
\indent if(~(rst\_ni \&\& ~jtag\_unlock \&\& ~rst\_9)) begin\\
\indent\indent for (j=0; j < 6; j=j+1) begin\\
\indent\indent\indent reglk\_mem[j] <= 'h0;\\
"""\\
\color{blue}
Register locks help prevent SoC peripherals' registers from malicious use of resources. The registers that can potentially leak secret data are locked by register locks. In the vulnerable code, the reglk\_mem is used for locking information. If one of its bits toggle to 1, the corresponding peripheral's registers will be locked.
A critical issue arises within the reset controller module. Specifically, the reset controller can inadvertently transmit a peripheral reset signal to the register lock within the user privilege domain.This unintentional action can result in the reset of the register locks, potentially exposing private data from all other peripherals, rendering them accessible and readable.
In this example the lock signal is reglk\_mem and the correct conditions for changing lock signals are (rst\_ni \&\& ~jtag\_unlock).\\
\color{black}
\noindent\rule{\linewidth}{0.4pt}

\end{document}